\documentstyle[12pt]{article}
\makeatletter
\newcommand{\singlespacing}{\let\CS=\@currsize\renewcommand{\baselinestretch}{1.0}\tiny\CS}
\newcommand{\doublespacing}{\let\CS=\@currsize\renewcommand{\baselinestretch}{1.5}\tiny\CS}

\begin{document}

\title{SWKB Quantization Rules for Bound States in Quantum Wells}

\vspace{2cm}

\author{Anjana Sinha\thanks{e-mail:res9523@www.isical.ac.in} \\
and \\
Rajkumar Roychoudhury\thanks{e-mail:raj@www.isical.ac.in}  \\
{\it Physics \& Applied Mathematics Unit}\\ 
{\it Indian Statistical Institute} \\ {\it Calcutta - 700035}\\ {\it India}}

\date{}

\maketitle

\thispagestyle{empty}

\setlength{\baselineskip}{19.5pt}

\centerline{\bf Abstract}

\vspace{0.3cm}

In a recent paper by M A F Gomes and S K Adhikari(J.Phys.(\bf B30) ,5987.(1997)), a matrix
formulation of the Bohr-Sommerfeld (mBS) quantization rule has been 
applied to the study of bound states in one-dimensional quantum wells.
They have observed that the usual Bohr-Sommerfeld (BS) and the
Wentzel-Kramers-Brillouin (WKB) quantization rules give poor estimates of
the eigen energies of the two confined trigonometric potentials, {\it viz.},
$ ~~~ V(x) ~=~ V_0 ~ cot~ ^2 ~~ \frac{\pi ~ x}{L} ~~~ $, 
and  the famous P\"{o}schl-Teller potential, \\
$~ ~~V(x) ~=~ V_{01} ~ cosec~ ^2 ~~ \frac{\pi ~ x}{2L} ~+~ V_{02} ~ 
sec~^ 2 ~~ \frac{\pi ~x}{2L}~~~ $, the WKB approach being  worse 
of the two, particularly for small values of $n$.
They suggested a matrix formulation of the Bohr-Sommerfeld method (mBS).
Though this technique improves the earlier results, it is not very  
accurate either.  Here we study these potentials in the framework 
of supersymmetric Wentzel-Kramers-Brillouin (SWKB)
approximation, and find that the SWKB quantization rule is superior 
to each one of the BS, mBS, and WKB approximations, 
as it reproduces the exact analytical results for the eigen energies.
Its added advantage is that it gives the correct analytical ground 
state wave functions as well.

\newpage

\noindent
{\bf Introduction :} 

The study of confined quantum systems is of considerable importance
in modern times as spatial confinement
significantly alters the physical and chemical properties of the 
system [1-4]. It influences the bond formation and 
chemical reactivity inside 
the cavities to a great extent. Even the optical properties (absorption and
emission of light in the visible or far infra-red range, Raman scattering) 
and electrical properties (capacitance and transport studies) change
radically. Hence this branch of Science is extremely useful in the study of
thermodynamic properties of non-ideal gases, investigation of atomic effects
in solids, in atoms and molecules under high pressure, impurity binding energy
in quantum wells, and even in the context of partially ionised plasmas.

      Various authors have empployed different techniques to study such
systems. Fairly recently, M A F Gomes and S K Adhikari [5] have suggested 
a matrix formulation of the Bohr-Sommerfeld ( mBS) quantization rule to 
give an estimate of the eigen energies of the Schr\"{o}dinger equation, 
for various one-dimensional quantum wells. They have compared the energies
thus obtained with those by Wentzel-Kramers-Brillouin (WKB) and usual 
Bohr-Sommerfeld ( BS) methods, as well as the exact analytical solution of the
Schr\"{o}dinger equation. They observed that in many cases the mBS
quantization rule yields more precise energies than the WKB or BS 
quantization rules. For small $n$ particularly, the WKB approximation gives
the poorest estimates.

         In this short comment, we study spatial 
confinement in the framework of SWKB (supersymmetric 
version of WKB) approximation. 
The motivation for the SWKB approach arises from the fact 
that this gives exact results in case of shape-invariant potentials. 
In this work, we deal with two trigonometric potentials, discussed in 
ref. [5], {\it viz.}, 
\begin{equation}
 V(x) ~=~ V_0 ~ cot~ ^2 ~~ \frac{\pi ~ x}{L} 
\end{equation}
and  the famous P\"{o}schl-Teller potential,
\begin{equation}
 V(x) ~=~ V_{01} ~ cosec~ ^2 ~~ \frac{\pi ~ x}{2L} ~+~ V_{02} ~ 
sec~^ 2 ~~ \frac{\pi ~x}{2L} 
\end{equation}
Both the potentials are tangentially limited by infinite walls 
at $x=0$ and $x=L$, ( $L$ being the dimension of the confining box ) 
and are of tremendous importance in molecular spectrocopy.
We find that our SWKB energies are identical to the exact 
analytical results [6,7] in both the cases. 
It is worth noting here that though the first potential is a special 
case of the second one with the identification $V_{01} = V_{02} 
= V_0 / 4 $ , it has been discussed by various authors due to its 
importance in molecular physics. Potential (1) represents a well 
symmetric around $ x=L/2$. Unless $V_{01}=V_{02}$, potential (2) represents 
an asymmetric well . For small $V_0, V_{01}, V_{02}, $ both the potentials 
represents perturbations on an infinite sqare well. Though both the
potentials are periodic in nature, the barriers put by the singularities
between the holes are impenetrable. So we consider a single hole only.  
The added advantage of the SWKB approximation is that it gives the exact
analytical ground state wave functions as well.

(Units used throughout are $\hbar ~=~ 2m~=~1 $. )

\vspace{1cm}

\noindent
{\bf Theory :}

First we give a brief outline of the SWKB method, starting 
from the ordinary WKB approximation. Writing the
potential $V(x)$ in terms of the superpotential $W(x)$ [8]
\begin{equation}
V(x) ~=~ W~^2 ~+~ W~^{\prime} ~(x) 
\end{equation}
the WKB quantization condition , viz.,
\begin{equation}
\displaystyle{\int ^b _a [E ~ - ~V(x) ] ~^{1/2} ~ dx ~=~ (n ~+~ 1/2 ) \pi 
\qquad n=0,1,2,...}
\end{equation}
where $a$, $b$ are the roots of the equation
\begin{equation}
E ~-~ V(x) ~=~ 0
\end{equation}
gets modified to the SWKB quantization condition
\begin{equation}
\displaystyle{\int ^d _c [E^{~\prime} ~ - ~W^2(x) ] ~^{1/2} ~ dx 
~-~ \frac{1}{2} \int ^d _c \frac{W~^{\prime}}{[E^{~\prime} ~-~ 
W^2 (x) ]^{1/2} } dx ~=~ (n ~+~ 1/2 ) \pi  \qquad n=0,1,2,...}
\end{equation}
where $c$, $d$ are the roots of the equation
\begin{equation}
E^{~\prime} ~-~ W^2(x) ~=~ 0
\end{equation}
Since the second integral has the value ${\pi /2} $, the SWKB quantization
rule reads
\begin{equation}
\displaystyle{\int ^d _c [E^{~\prime} ~ - ~W^2(x) ] ~^{1/2} ~ dx  
~=~ n ~ \pi  \qquad n=1,2,3,...}
\end{equation}
The ground state 
\begin{equation}
\psi _0 = exp ~( \int ~ W(x)~ dx)
\end{equation}
will be normalizable if $ \int ~ W(x) ~ dx $ exists.

\vspace{1cm}

   For the P\"{o}schl-Teller potential 
\begin{equation}
V(x) ~=~ V_{01} ~ cosec~ ^2 ~~ \frac{\pi ~ x}{2L} ~+~ V_{02} ~ 
sec~^ 2 ~~ \frac{\pi ~x}{2L}
\end{equation}
the superpotential can be taken as 
\begin{equation}
W(x) ~=~ A_1 ~ cot ~ \alpha x ~+~ A_2 ~ tan ~ \alpha x 
\end{equation}
Now the Schr\"{o}dinger equation 
\begin{equation}
\left( ~ - ~ \frac{d^2}{dx^2} ~+~ V(x)~ \right) ~ \psi ~=~ E~ \psi
\end{equation}
can be written in the form
\begin{equation}
\left( ~ - ~ \frac{d^2}{dx^2} ~+~ (W^{~2} ~+~ W^{~\prime} )
~ \right) ~ \psi ~=~ E^{~\prime} ~ \psi
\end{equation} 
with
\begin{equation}
\alpha  ~=~ \frac{\pi}{2L}
\end{equation}
\begin{equation}
V_{01} ~=~ A_1 ^2 ~-~ A_1 \alpha
\end{equation}
\begin{equation}
V_{02} ~=~ A_2 ^2 ~+~ A_2 \alpha
\end{equation}
\begin{equation}
E^{~\prime} ~=~ E ~-~ ( A_1 ~-~ A_2 ) ^2
\end{equation}
Hence the SWKB condition (8) takes the form
\begin{equation}
\displaystyle{  ~ \int ^{x_2} _{x_1} \left\{ 
E^{~\prime} ~-~ {A_1 ^2}~{cosec ^{~2} {\alpha x}} ~-~ {A_2 ^2}~{ sec ^{~2}
{\alpha x}} ~+~ (A_1 ~-~ A_2 )^2 \right\}^{~1/2} ~dx
 ~=~ n ~ \pi  \qquad n=0,1,2,...}
\end{equation}
which can be written as
\begin{equation}
\displaystyle{ \frac{1}{ \alpha} ~ \int ^{z_2} _{z_1} \left\{ 
E ~-~ \frac{A_1 ^2}{z ^2 } ~-~ \frac{A_2 ^2}{ 1~-~z^2}
\right\}^{~1/2} ~\frac{dz}{(1~-~z^2)^{~1/2}}
 ~=~ n ~ \pi  \qquad n=0,1,2,...}
\end{equation}
where 
\begin{equation}
z~=~ sin~( \alpha x)
\end{equation}
Putting 
\begin{equation}
z^2~=~t
\end{equation}
(19) reduces to
\begin{equation}
\displaystyle{ \frac{1}{2 \alpha} ~ \int ^{t_2} _{t_1} \left\{ 
~ \frac{ -~A_1 ^2 ~+~ (A_1 ^2 ~-~ A_2 ^2 ~+~ E) t ~-~ Et^2}{ t ( 1~-~ t)} ~
\right\} ~ dt 
 ~=~ n ~ \pi  \qquad n=0,1,2,...}
\end{equation}
This can be evaluated with the help of formulae given in [9].

Omitting the detailed calculations for brevity,
we quote the SWKB energy directly.
\begin{equation}
E_n ^{swkb}= [2 ~\alpha ~n ~+~ (A_1 ~-~ A_2 )]^2 \qquad n~=~0,1,2,...
\end{equation}
Writing
$$ V_{01}~=~v_1 ~ E_1 ^{\infty} $$
$$ V_{02}~=~v_2 ~ E_1 ^{\infty} $$
$$ E_n~=~\epsilon _n ~ E_1 ^{\infty} $$
the SWKB energy can be reformulated as 
\begin{equation}
E_p ~=~ E^{~ \infty} _1 ~ \left \{ p ~+~ \frac{1}{2} ~+~ \sqrt{ v_{01} 
~+~ \frac{1}{16} } ~+~ \sqrt{ v_{02} ~+~ \frac{1}{16}} \right \} 
\qquad p ~=~ 1,2,3,...
\end{equation}
which is identical to the exact analytical formula for the Schr\"{o}dinger
eigenenergies  [6]. Thus our SWKB approach reproduces the exact eigenenergies
of the P\"{o}schl-Teller potential.

Also the ground state eigenfunction $ \psi _0  ~=~ |N_0| ~ 
exp( \int W_0 ~dx)$ takes the form
\begin{equation}
\psi _0 ~=~ |N_0| ~ \frac{sin^{~A_1 / \alpha} ~ \alpha x}{
cos^{~A_2/ \alpha} ~ \alpha x}
\end{equation}
where $ N_0 $ is fixed by normalization.
For $ \psi ~=~ 0 $ at $x=0$ and $x=L$, $A_1$ and $A_2$ must satisfy the
condition 
$$ A_1 ~>~ 0  ~~~ , ~~~ A_2 ~<~ 0 $$
Thus the SWKB approach reproduces the exact ground state wave function 
of the system [6].

\vspace{1cm}

For the special case
$$ V_{01} ~=~ V_{02} ~=~ V_0 / 4 $$
the P\"{o}schl-Teller potential may be cast in the form
\begin{equation}
V(x) ~=~ V_0 ~ cot~ ^2 ~~ \frac{\pi ~ x}{L}  
\end{equation}
The superpotential can be taken to be 
\begin{equation}
W(x)~=~A~cot~ \alpha x 
\end{equation}
with
\begin{equation}
\alpha  ~=~ \frac{\pi}{L}
\end{equation}
\begin{equation}
V_0 ~=~ A ^2 ~-~ A \alpha
\end{equation}
\begin{equation}
E ~=~ E^{~\prime} ~+~  A \alpha
\end{equation}
so that the SWKB quantization condition (8) gives 
\begin{equation}
\displaystyle{\int ^d _c [E^{~\prime} ~ - A^{~2}~cot^{~2} 
( \frac{ \pi x}{L}) ] ~^{1/2}  ~ dx  
~=~ n ~ \pi  \qquad n=0,1,2,...}
\end{equation}
Putting
\begin{equation}
z=A~cot~( \frac{ \pi x}{L})
\end{equation}
(31) can be written as 
\begin{equation}
\displaystyle{ - ~ \frac{L}{A \pi} ~ \int ^{z_2} _{z_1} \frac{ [E ~ - z^{~2}]
 ~^{1/2}}{1~+~ \frac{z^2}{A^2}} ~dz
 ~=~ n ~ \pi  \qquad n=0,1,2,...}
\end{equation}
Further substituting
\begin{equation}
\rho ~=~ \frac{z}{(E~-~ z^2)^{~1/2}}
\end{equation}
(33) reduces to, after some algebra,
\begin{equation}
\displaystyle{ \frac{2AL}{ \pi} ~ \int ^{\infty} _{0} \left\{ 
\frac{1}{1~+~ \rho ^2 } ~-~ \frac{E~+~A^2}{ \rho^2 ~(E~+~A^2) ~+~ A^2}
\right\} ~d \rho
 ~=~ n ~ \pi  \qquad n=0,1,2,...}
\end{equation}
Once again omitting the lengthy calculations,
it is found that the energy turns out to be
\begin{equation}
E^{~swkb} _n ~=~ \frac{\pi ^2 }{2m ~ L^2} ~ \left \{ \sqrt{ \frac{2m ~ L^2 ~
A^2}{ \pi ^2}} ~+~ n \right \} ^2 ~-~ A^2 ~+~ \frac{A \pi}{L} ~~ \qquad ~~
n ~= 0,1,2,3,.....
\end{equation}
Using eqns. (28) and (29), eqn.(36) can be cast in the form of the 
exact Schr\"{o}dinger energy for the potential
under consideration [7] 
\begin{equation}
E_p ~=~ E^{~ \infty} _1 \left \{ p^2 ~+~ (p ~-~ \frac{1}{2}) ~ 
\sqrt{4v ~+~ 1} ~-1 \right \} ~~~ \qquad ~~~ p= 1,2,3,.....
\end{equation}
with the identification
\begin{equation}
V_0 ~=~ v ~ E^{~ \infty} _1  
\end{equation}
\begin{equation}
E^{\infty} _n ~=~ \frac{\pi ^2 ~ n^2}{~ L^2}
\end{equation}

Similarly the ground state wave function 
$$ \psi _0  ~=~ |c_0| ~ exp( \int W_0 ~dx)$$
\begin{equation}
\qquad ~~~~~~~~~ =~|c_0| ~ sin ^{~A / \alpha} ~ \alpha x
\end{equation}
coincides with the exact formula for the system [7], 
where $ c_0 $ is the normalization factor.
For $ \psi ~=~ 0 $ at $x=0$ and $x=L$, $A$ must satisfy the
condition 
$$ A ~>~ 0  $$

\vspace{1cm}

\noindent
{\bf Conclusions :}

 In this short comment, we have studied  spatial 
confinement in the framework of SWKB (supersymmetric 
version of WKB) approximation. 
In particular, we have dealt with two trigonometric potentials, discussed in 
ref. [5], {\it viz.} , 
\begin{equation}
 V(x) ~=~ V_0 ~ cot~ ^2 ~~ \frac{\pi ~ x}{L} 
\end{equation}
and  the famous P\"{o}schl-Teller potential,
\begin{equation}
 V(x) ~=~ V_{01} ~ cosec~ ^2 ~~ \frac{\pi ~ x}{2L} ~+~ V_{02} ~ 
sec~^ 2 ~~ \frac{\pi ~x}{2L} 
\end{equation}
Both the potentials are tangentially limited by infinite walls 
at $x=0$ and $x=L$, ( $L$ being the dimension of the confining box ) 
and are of tremendous importance in molecular spectrocopy.
 It had been observed in ref. [5]
that mBS eigenenergies are somewhat better than BS and / or WKB ones , the WKB
approximation giving the worst results for small values of $n$. 
We find that our SWKB quantization rule is 
far better than each one of the BS,
mBS, and WKB approximations , as it reproduces the exact analytical 
eigenenergies for both the potentials. 
Also in the SWKB approach, we obtain the exact
analytical ground state eigenfunction.

\newpage

\noindent
{\bf Acknowledgment :}

 One of the authors (A.S.) acknowledges financial assistance from the 
Council of Scientific and Industrial Research, India.

\vspace{1cm}

\noindent
{\bf References :}

\begin{enumerate}

\item[[1]] C. Zicovich-Wilson, W. Jask\'{o}lski and J.H. Planelles. Int. J.
Quantum Chem. (1995) {\bf 54} 61-72.
\item[[2]] C. Zicovich-Wilson, W. Jask\'{o}lski and J.H. Planelles. Int. J.
Quantum Chem. (1994) {\bf 50} 429-444.
\item[[3]] S.A. Cruz, E Ley-Koo, J.L. Marin and A. Taylor Armitage. Int. J.
Quantum  Chem. (1995) {\bf 54} 3-11.
\item[[4]] C. Zicovich-Wilson, A. Corma and P. Viruela. J. Phys. Chem.
(1994) {\bf 98} 10863-10870. 
\item[[5]] M.A.F Gomes and S.K. Adhikari.  J. Phys. B : At. Mol. Opt. Phys.
(1997)  {\bf 30} 5987-5997.
\item[[6]] S. Flugge. Practical Quantum Mechanics I (Berlin: Springer, 1976).
\item[[7]] I. I. Gol'dman and V. D. Krivchenkov. {\bf Problems in Quantum
Mechanics.} {\it Dover Publications, Inc., New York} (1993).
\item[[8]] A. Sinha, R. Roychoudhury and Y. P. Varshni. Can. J. Phys.
(1996) {\bf 74}  39-42. 
\item[[9]] I. S. Gradshteyn and I. M. Ryzhik. {\bf Tables of Integrals, 
Series and Products.} {\it Academic Press} (1992).
	
\end{enumerate}
\end{document}